\documentclass[12pt]{article}
\usepackage{a4wide}
\usepackage{latexsym}
\usepackage{cite}

\usepackage{pslatex}
\usepackage[latin1]{inputenc}
\usepackage[T1]{fontenc}

\def\bq{\begin{eqnarray}}
\def\eq{\end{eqnarray}}
\def\l{\langle}
\def\r{\rangle} 
\def\eps{\varepsilon}

\begin{document}

\thispagestyle{empty}

\begin{flushright}
  MZ-TH/06-10
\end{flushright}

\vspace{1.5cm}

\begin{center}
  {\Large\bf NNLO corrections to 2-jet observables in electron-positron annihilation\\
  }
  \vspace{1cm}
  {\large Stefan Weinzierl\\
\vspace{2mm}
      {\small \em Institut f{\"u}r Physik, Universit{\"a}t Mainz,}\\
      {\small \em D - 55099 Mainz, Germany}\\
  } 
\end{center}

\vspace{2cm}

\begin{abstract}\noindent
  {
I report on a numerical program, which can be used to calculate any infra-red safe two-jet
observable in electron-positron annihilation to next-to-next-to-leading order 
in the strong coupling constant $\alpha_s$.
The calculation is based on the subtraction method.
The result for the two-jet cross section is compared to the literature.
   }
\end{abstract}

\vspace*{\fill}

\newpage

\section{Introduction}
\label{sec:intro}

The forthcoming LHC experiment will provide a large sample of multi-particle
final states.
In order to extract information from this data, precise theoretical calculations
are necessary.
This implies to extend perturbative calculations for selected processes
from next-to-leading order (NLO)
to next-to-next-to-leading order (NNLO) in the perturbative expansion in the strong
coupling constant.
Due to a large variety of interesting jet observables it is desirable not to perform
this calculation for a specific observable, but to set up a computer
program, which yields predictions for any infra-red safe observable relevant to the process
under consideration.
Such a task requires the calculation of two-loop amplitudes,
a method for the cancellation of infrared divergences and stable and efficient Monte
Carlo techniques.

The past years have witnessed a tremendous progress in techniques 
for the computation of two-loop integrals
\cite{Tarasov:1996br,Smirnov:1999gc,Tausk:1999vh,Laporta:2000dc,Gehrmann:1999as,Moch:2001zr,Weinzierl:2002hv,Moch:2005uc,Czakon:2005rk,Anastasiou:2004vj}
and in the calculation of two-loop amplitudes 
\cite{Bern:2000ie,Bern:2000dn,Anastasiou:2000kg,Glover:2001af,Bern:2001dg,Bern:2001df,Garland:2001tf,Moch:2002hm}.
In addition, several options for the cancellation of infrared divergences have been discussed
\cite{Kosower:2002su,Weinzierl:2003fx,Anastasiou:2003gr,Gehrmann-DeRidder:2003bm,Gehrmann-DeRidder:2004tv,Gehrmann-DeRidder:2005cm,Binoth:2004jv,Heinrich:2006sw,Kilgore:2004ty,Frixione:2004is,Somogyi:2005xz}.
Among those, the subtraction method -- well-known from NLO computations
\cite{Frixione:1996ms,Catani:1997vz,Dittmaier:1999mb,Phaf:2001gc,Catani:2002hc} --
and sector decomposition 
\cite{Hepp:1966eg,Roth:1996pd,Binoth:2000ps}
are the most promising candidates.
For the method based on sector decomposition first numerical results have become available
for the processes $e^+ e^- \rightarrow \mbox{2 jets}$ and $H \rightarrow \gamma \gamma$
\cite{Anastasiou:2004qd}.
Up to now, no numerical NNLO program which is based on the subtraction method and which allows
arbitrary cuts and measurement functions is available.

In this paper I consider the process $e^+ e^- \rightarrow \mbox{2 jets}$ and
I report on a numerical program, which can be used to calculate 
any infra-red safe observable related to this process to next-to-next-to-leading order.
The calculation is based on the subtraction method.
The purpose of this article is to demonstrate the feasability and correctness of the subtraction
method at NNLO in a non-trivial example.
In the set-up of the program nothing is specific to the process
$e^+ e^- \rightarrow \mbox{2 jets}$ -- except the matrix elements and the fact that in 
electron-positron annihilation infra-red singularities occur only in the final state.
Therefore this process -- apart from being of interest by itself -- serves also as a test-ground
for the subtraction method for other processes like
$e^+ e^- \rightarrow \mbox{3 jets}$ or 
$p p \rightarrow \mbox{2 jets}$.

I will devote particular attention to the way the phase space is sampled in the Monte-Carlo integration
and describe methods employed to reduce the statistical error of the Monte-Carlo integration.

This paper is organized as follows:
In the next section the subtraction method is reviewed.
Section~\ref{sec:phasespace} describes techniques for an efficient generation of the phase space.
In section~\ref{sec:num} the numerical results are presented.
Finally, the conclusions are contained in section~\ref{sec:concl}.
In an appendix, I collected the explicit formulae for the subtraction terms 
for double unresolved configurations.

\section{The subtraction method}
\label{sec:conv}

The master formula to calculate an observable at an collider with no 
initial-state hadrons (e.g. an electron-positron collider) is given by
\bq
\label{master_formula}
\l {\cal O}^{(j)} \r & = & \frac{1}{2 K(s)}
             \frac{1}{\left( 2 J_1+1 \right)}
             \frac{1}{\left( 2 J_2+1 \right)}
             \sum\limits_n
             \int d\phi_{n-2}
             {\cal O}^{(j)}_n\left(p_1,...,p_n\right)
             \sum\limits_{helicity} 
             \left| {\cal A}_n \right|^2
\eq
where $p_1$ and $p_2$ are the momenta of the initial-state particles, 
$2K(s)=2s$ is the flux factor and $s=(p_1+p_2)^2$ is the center-of-mass energy squared.
The factors $1/(2J_1+1)$ and $1/(2J_2+1)$ correspond
to an averaging over the initial helicities. $d\phi_{n-2}$ is the invariant phase space measure
for $(n-2)$ final state particles and
${\cal O}^{(j)}_n\left(p_1,...,p_n\right)$ is the observable, evaluated with an $n$-parton configuration.
The index $j$ indicates that the leading order contribution depends on $j$ partons.
The observable has to be infra-red safe, in particular this implies that in single and double
unresolved limits we must have
\bq
{\cal O}^{(j)}_{n+1}(p_1,...,p_{n+1}) & \rightarrow & {\cal O}^{(j)}_n(p_1',...,p_n')
 \;\;\;\;\;\;\mbox{for single unresolved limits},
 \nonumber \\
{\cal O}^{(j)}_{n+2}(p_1,...,p_{n+2}) & \rightarrow & {\cal O}^{(j)}_n(p_1',...,p_n')
 \;\;\;\;\;\;\mbox{for double unresolved limits}.
\eq
${\cal A}_n$ is the amplitude with $n$ partons.
At NNLO we need the following expansions of the amplitudes:
\bq
& & 
\hspace*{-5mm}
 \left| {\cal A}_n \right|^2
 =  
   \left. {\cal A}_n^{(0)} \right.^\ast {\cal A}_n^{(0)} 
 + 
   \left(
             \left. {\cal A}_n^{(0)} \right.^\ast {\cal A}_n^{(1)} 
           + \left. {\cal A}_n^{(1)} \right.^\ast {\cal A}_n^{(0)} 
       \right)
 + 
   \left(
             \left. {\cal A}_n^{(0)} \right.^\ast {\cal A}_n^{(2)} 
           + \left. {\cal A}_n^{(2)} \right.^\ast {\cal A}_n^{(0)}  
           + \left. {\cal A}_n^{(1)} \right.^\ast {\cal A}_n^{(1)} 
       \right),
 \nonumber \\
& &
\hspace*{-5mm}
  \left| {\cal A}_{n+1} \right|^2
 =  
   \left. {\cal A}_{n+1}^{(0)} \right.^\ast {\cal A}_{n+1}^{(0)}
 + 
   \left(
          \left. {\cal A}_{n+1}^{(0)} \right.^\ast {\cal A}_{n+1}^{(1)} 
        + \left. {\cal A}_{n+1}^{(1)} \right.^\ast {\cal A}_{n+1}^{(0)}
 \right),
 \nonumber \\ 
& &
\hspace*{-5mm}
  \left| {\cal A}_{n+2} \right|^2
 = 
   \left. {\cal A}_{n+2}^{(0)} \right.^\ast {\cal A}_{n+2}^{(0)}. 
\eq
Here ${\cal A}_n^{(l)}$ denotes an amplitude with $n$ partons and $l$ loops.
We rewrite the master formula eq.~(\ref{master_formula})
symbolically as 
\bq
\l {\cal O}^{(j)} \r & = & 
             \sum\limits_n \int {\cal O}^{(j)}_{n} \; d\sigma_{n}
\eq
and the LO, NLO and NNLO contribution as 
\bq
\label{def_LO_NLO_NNLO}
\l {\cal O}^{(j)} \r^{LO} & = & \int {\cal O}^{(j)}_{n} \; d\sigma_{n}^{(0)},
 \nonumber \\
\l {\cal O}^{(j)} \r^{NLO} & = & \int {\cal O}^{(j)}_{n+1} \; d\sigma_{n+1}^{(0)} + \int {\cal O}^{(j)}_{n} \; d\sigma_{n}^{(1)},
 \nonumber \\
\l {\cal O}^{(j)} \r^{NNLO} & = & \int {\cal O}^{(j)}_{n+2} \; d\sigma_{n+2}^{(0)} 
                   + \int {\cal O}^{(j)}_{n+1} \; d\sigma_{n+1}^{(1)} 
                   + \int {\cal O}^{(j)}_{n} \; d\sigma_{n}^{(2)}.
\eq
The individual contributions on the r.h.s. of eq.~(\ref{def_LO_NLO_NNLO})
to $\l {\cal O}^{(j)} \r^{NLO}$ and $\l {\cal O}^{(j)} \r^{NNLO}$ are in general infra-red divergent, only the sum is finite.
However, these contributions live on different phase spaces, which prevents a naive Monte Carlo approach.
To render the individual contributions finite, one adds and subtracts suitable chosen terms.
The NLO contribution is given by
\bq
\l {\cal O}^{(j)} \r^{NLO} & = & 
   \int \left( {\cal O}^{(j)}_{n+1} \; d\sigma_{n+1}^{(0)} - {\cal O}^{(j)}_{n} \circ d\alpha^{(0,1)}_{n} \right)
 + \int \left( {\cal O}^{(j)}_{n} \; d\sigma_{n}^{(1)} + {\cal O}^{(j)}_{n} \circ d\alpha^{(0,1)}_{n} \right).
\eq
The notation ${\cal O}^{(j)}_{n} \circ d\alpha^{(0,1)}_{n}$ is a reminder, that
in general the approximation is a sum of terms
\bq
{\cal O}^{(j)}_{n} \circ d\alpha^{(0,1)}_{n} & = & \sum {\cal O}^{(j)}_{n} \; d\alpha^{(0,1)}_{n}
\eq
and the mapping used to relate the $n+1$ parton configuration to a $n$ parton configuration
differs in general for each summand.
\\
\\
In a similar way, the NNLO contribution is written as
\bq
\l {\cal O}^{(j)} \r^{NNLO} & = &
 \int \left( {\cal O}^{(j)}_{n+2} \; d\sigma_{n+2}^{(0)} 
             - {\cal O}^{(j)}_{n+1} \circ d\alpha^{(0,1)}_{n+1}
             - {\cal O}^{(j)}_{n} \circ d\alpha^{(0,2)}_{n} 
      \right) \nonumber \\
& &
 + \int \left( {\cal O}^{(j)}_{n+1} \; d\sigma_{n+1}^{(1)} 
               + {\cal O}^{(j)}_{n+1} \circ d\alpha^{(0,1)}_{n+1}
               - {\cal O}^{(j)}_{n} \circ d\alpha^{(1,1)}_{n}
        \right) \nonumber \\
& & 
 + \int \left( {\cal O}^{(j)}_{n} \; d\sigma_n^{(2)} 
               + {\cal O}^{(j)}_{n} \circ d\alpha^{(0,2)}_{n}
               + {\cal O}^{(j)}_{n} \circ d\alpha^{(1,1)}_{n}
        \right).
\eq
$d\alpha^{(0,1)}_{n+1}$ is the NLO subtraction term for $(n+1)$-parton configurations,
$d\alpha^{(0,2)}_{n}$ and $d\alpha^{(1,1)}_{n}$ are generic NNLO subtraction terms.
It is convenient to split these terms into
\bq
 d\alpha^{(0,2)}_{n} & = & d\alpha^{(0,2)}_{(0,0),n} - d\alpha^{(0,2)}_{(0,1),n},
 \nonumber \\
 d\alpha^{(1,1)}_{n} & = & d\alpha^{(1,1)}_{(1,0),n} + d\alpha^{(1,1)}_{(0,1),n},
\eq
such that $d\alpha^{(0,2)}_{(0,0),n}$ and $d\alpha^{(1,1)}_{(1,0),n}$ approximate 
$d\sigma_{n+2}^{(0)}$ and $d\sigma_{n+1}^{(1)}$, respectively.
$d\alpha^{(0,2)}_{(0,1),n}$ and $d\alpha^{(1,1)}_{(0,1),n}$ are approximations to
$d\alpha^{(0,1)}_{n+1}$.

\subsection{The amplitudes and the subtraction terms}

The NNLO correction to $e^+ e^- \rightarrow \;\mbox{2 jets}$ requires the amplitudes for
$e^+ e^- \rightarrow q \bar{q}$ up to two-loops, the amplitudes
$e^+ e^- \rightarrow q g \bar{q}$ up to one-loop, and the Born amplitudes
for $e^+ e^- \rightarrow q g g \bar{q}$ and 
$e^+ e^- \rightarrow q \bar{q} q' \bar{q}'$.
All these amplitudes are known and can be found in the 
literature
\cite{Matsuura:1988wt,Matsuura:1989sm,Kramer:1987sg,Ellis:1981wv,Schuler:1987ej,Korner:1990sj,Giele:1992vf,Ali:1979rz}.
I will label the momenta of the particles of the amplitudes as follows:
\bq
 {\cal A}_2^{(j)}(1,2) & = & {\cal A}_2^{(j)}(q_1,\bar{q}_2,e^+_3,e^-_4),
 \nonumber \\
 {\cal A}_3^{(j)}(1,2,3) & = & {\cal A}_3^{(j)}(q_1,g_2,\bar{q}_3,e^+_4,e^-_5),
 \nonumber \\
 {\cal A}_{4,qgg\bar{q}}^{(j)}(1,2,3,4) & = & {\cal A}_{4,qgg\bar{q}}^{(j)}(q_1,g_2,g_3,\bar{q}_4,e^+_5,e^-_6),
 \nonumber \\
 {\cal A}_{4,q\bar{q}q'\bar{q}'}^{(j)}(1,2,3,4) & = & {\cal A}_{4,q\bar{q}q'\bar{q}'}^{(j)}(q_1,\bar{q}_2,q_3',\bar{q}_4',e^+_5,e^-_6).
\eq
For the subtraction terms I follow the approach of ref.~\cite{Gehrmann-DeRidder:2005cm}
and I use spin-averaged antenna functions.
Compared to dipole subtraction terms, the use of antenna subtraction terms leads to fewer subtraction terms
and therefore to a faster program.
I follow closely the notation of the authors in \cite{Gehrmann-DeRidder:2005cm}
and denote the three-parton antenna functions by 
\bq
A_3^{(l)}(q, g, \bar{q}), 
\;\;\;
D_3^{(l)}(q, g, g), 
\;\;\;
E_3^{(l)}(q, q', \bar{q}'), 
\eq
depending on which particles form the antenna. A subscript ``sc'' or ``nf'' is used to indicate contributions
sub-leading in colour or proportional to the number of light flavours $N_f$.
The four-parton antenna functions are denoted by
\bq
A_4^{(0)}(q, g, g, \bar{q}), 
 \;\;\;
B_4^{(0)}(q, q', \bar{q}', \bar{q}), 
 \;\;\;
C_4^{(0)}(q, q, \bar{q}, \bar{q}).
\eq
The NLO subtraction term is rather simple and given by
\bq
 d\alpha_2^{(0,1)} & = &
 \frac{N^2-1}{2 N} A_3^0(1,2,3) \left| {\cal A}_2^{(0)}(1',2') \right|^2.
\eq
The amplitude ${\cal A}_2^{(0)}$ is evaluated with momenta, which are obtained from the original momenta
as follows \cite{Kosower:1998zr}: If $(p_i,p_j,p_k)$ is a set of momenta corresponding to an antenna, 
such that particle $j$ is emitted by the antenna formed
by particles $i$ and $k$, then the mapped momenta are given by
\bq
\label{momenta_map}
p_I & = & \frac{(1+\rho)s_{ijk} - 2 r s_{jk}}{2(s_{ijk}-s_{jk})} p_i 
          + r p_j
          + \frac{(1-\rho)s_{ijk} - 2 r s_{ij}}{2(s_{ijk}-s_{ij})} p_k,
 \nonumber \\
p_K & = & \frac{(1-\rho)s_{ijk} - 2 (1-r) s_{jk}}{2(s_{ijk}-s_{jk})} p_i 
          + (1-r) p_j
          + \frac{(1+\rho)s_{ijk} - 2 (1-r) s_{ij}}{2(s_{ijk}-s_{ij})} p_k,
\eq
where
\bq
 r= \frac{s_{jk}}{s_{ij}+s_{jk}},
 & &
 \rho = \sqrt{1 + 4 r (1-r) \frac{s_{ij}s_{jk}}{s_{ijk} s_{ik}} }.
\eq
For the NNLO subtraction terms, one first needs the NLO subtraction terms for
$e^+ e^- \rightarrow \;\mbox{3 jets}$. These are given by
\bq
 d\alpha_{3,qgg\bar{q}}^{(0,1)} & = &
 \frac{1}{2} \left\{ 
  \frac{N}{2} \left[ D_3^0(1,2,3) + D_3^0(1,3,2) + D_3^0(4,2,3) + D_3^0(4,3,2) \right]
 \right.
 \nonumber \\
 & & \left.
  - \frac{1}{2N} \left[ A_3^0(1,2,4) + A_3^0(1,3,4) \right]
 \right\} \circ \left| {\cal A}_3^{(0)} \right|^2,
 \nonumber \\
 d\alpha_{3,q\bar{q}q'\bar{q}'}^{(0,1)} & = &
 \left\{
 \left( \frac{1}{4} + \frac{N_f-1}{2} \right)
  \frac{N}{2} \left[ E_3^0(1,3,4) + E_3^0(2,4,3) + E_3^0(3,1,2) + E_3^0(4,2,1) \right]
 \right. 
 \nonumber \\
 & &
 \left.
 + \frac{1}{4} 
  \frac{N}{2} \left[ E_3^0(1,3,2) + E_3^0(2,4,1) + E_3^0(3,1,4) + E_3^0(4,2,3) \right]
 \right\} \circ \left| {\cal A}_3^{(0)} \right|^2.
\eq
Again, the amplitude ${\cal A}_3^{(0)}$ is evaluated with momenta obtained from eq.~(\ref{momenta_map}),
which depend now on the partons, which form the antenna.
Integration over an one-parton phase space yields
\bq
\lefteqn{
 \int\limits_1 d\alpha_3^{(0,1)} = 
 \int\limits_1 \left( d\alpha_{3,qgg\bar{q}}^{(0,1)} + d\alpha_{3,q\bar{q}q'\bar{q}'}^{(0,1)} \right)
 = } & & \nonumber \\
 & &
 \left\{
  \frac{N}{2} \left[ {\cal D}_3^0(s_{12}) + N_f {\cal E}_3^0(s_{12}) 
                   + {\cal D}_3^0(s_{23}) + N_f {\cal E}_3^0(s_{23}) 
              \right]
  - \frac{1}{2N} {\cal A}_3^0(s_{13}) 
 \right\} \left| {\cal A}_3^{(0)} \right|^2.
\eq
The iterated subtraction terms for double unresolved contributions read
\bq
\lefteqn{
 d\alpha^{(0,2)}_{(0,1),2,qgg\bar{q}} = 
 \frac{1}{2} \left\{ 
  \frac{N}{2} \left[ D_3^0(1,2,3) + D_3^0(1,3,2) + D_3^0(4,2,3) + D_3^0(4,3,2) \right]
 \right.
 }
 \nonumber \\
 & & \left.
  - \frac{1}{2N} \left[ A_3^0(1,2,4) + A_3^0(1,3,4) \right]
 \right\} 
 \circ \frac{N^2-1}{2N}
 A_3^0(1',2',3') \left| {\cal A}_2^{(0)}(1'', 2'') \right|^2,
 \hspace*{40mm}
 \nonumber \\
\lefteqn{
 d\alpha^{(0,2)}_{(0,1),2,q\bar{q}q'\bar{q}'} = 
 \left\{
 \left( \frac{1}{4} + \frac{N_f-1}{2} \right)
  \frac{N}{2} \left[ E_3^0(1,3,4) + E_3^0(2,4,3) + E_3^0(3,1,2) + E_3^0(4,2,1) \right]
 \right. 
}
 \nonumber \\
 & &
 \left.
 + \frac{1}{4} 
  \frac{N}{2} \left[ E_3^0(1,3,2) + E_3^0(2,4,1) + E_3^0(3,1,4) + E_3^0(4,2,3) \right]
 \right\} 
 \nonumber \\
 & &
 \circ \frac{N^2-1}{2N}
 A_3^0(1',2',3') \left| {\cal A}_2^{(0)}(1'', 2'') \right|^2.
\eq
Here we first use the momentum mapping in eq.~(\ref{momenta_map})
to relate the four final-state partons to a three parton configuration $(1',2',3')$.
The momentum mapping is then used a second time to obtain the configuration $(1'', 2'')$.
The subtraction terms for double unresolved contributions read
\bq
\lefteqn{
 d\alpha^{(0,2)}_{(0,0),2,qgg\bar{q}} = 
 \frac{1}{2}
 \left\{
 \frac{N}{2} \frac{N^2-1}{2N} \left[ A_4^0(1,2,3,4) + A_4^0(1,3,2,4) \right]
 \right. 
 } & &
 \nonumber \\
 & & \left.
 - \frac{1}{2N} \frac{N^2-1}{2N} \left[ A_{4,sc}^0(1,2,3,4) + A_{4,sc}^0(1,3,2,4) \right]
 \right\}
 \circ \left| {\cal A}_2^{(0)} \right|^2,
 \nonumber \\
\lefteqn{
 d\alpha^{(0,2)}_{(0,0),2,q\bar{q}q'\bar{q}'} = } & &
 \nonumber \\
 & &
 \left\{
 \frac{C_F}{2} \left[ \left( \frac{1}{4} + \frac{N_f-1}{2} \right)
                        \left( B_4^0(2,4,3,1) + B_4^0(4,2,1,3) \right)
                        + \frac{1}{4} \left( B_4^0(2,4,1,3) + B_4^0(4,2,3,1) \right) \right]
 \right. 
 \nonumber \\
 & & \left.
 - \frac{C_F}{2N} \left[ C_4^0(2,4,3,1) + C_4^0(4,2,1,3) + C_4^0(2,4,1,3) + C_4^0(4,2,3,1) \right]
 \right\}
 \circ \left| {\cal A}_2^{(0)} \right|^2.
\eq
The amplitude ${\cal A}_2^{(0)}$ is again evaluated with a two-parton final-state configuration, obtained
through iteration of eq.~(\ref{momenta_map}). For the cyclic order $(i,j,k,l)$ one first compares $s_{ij}$ to $s_{kl}$.
If $s_{ij}$ is smaller, one first combines $p_i, p_j, p_k$ into $p_I', p_K'$ and then in a second step
$p_I',p_K',p_l$ into $p_I'',p_L''$.
The case $s_{kl}<s_{ij}$ is analogous.
Finally we need the subtraction terms for one-loop amplitudes with one unresolved parton, They are given 
by
\bq
 d\alpha^{(1,1)}_{(1,0),2} & = &
 \frac{N^2-1}{2 N} A_3^0(1,2,3) \left| {\cal A}_2^{(1)} \right|^2
 \nonumber \\
 & &
 + \frac{N^2-1}{4} \left[ A_3^1(1,2,3) + \frac{N_f}{N} A_{3,nf}^1(1,2,3) - \frac{1}{N^2} A_{3,sc}^1(1,2,3) \right]
 \left| {\cal A}_2^{(1)} \right|^2,
 \nonumber \\
 d\alpha^{(1,1)}_{(0,1),2} & = &
 \left\{
  \frac{N}{2} \left[ {\cal D}_3^0(s_{12}) + N_f {\cal E}_3^0(s_{12}) + {\cal D}_3^0(s_{23}) + N_f {\cal E}_3^0(s_{23}) \right]
  - \frac{1}{2N} {\cal A}_3^0(s_{13}) 
 \right\} 
 \nonumber \\
 & &
 \frac{N^2-1}{2N}
 A_3^0(1,2,3) \left| {\cal A}_2^{(0)} \right|^2.
\eq
The cancellation of explicit poles of the dimensional regularization parameter $\eps$ occurs individually in the following
combintations:
\bq
 d\sigma_{n+1}^{(0)} + \int\limits_1 d\alpha_{(n+1)}^{(0,1)} 
 & = & {\cal O}\left(\eps^0\right),
 \nonumber \\
 d\alpha_{(1,0)}^{(1,1)} + d\alpha_{(0,1)}^{(1,1)} & = & {\cal O}\left(\eps^0\right),
 \nonumber \\
 d\sigma_n^{(2)} + \int\limits_1 d\alpha^{(1,1)} + \int\limits_2 d\alpha^{(0,2)} & = & {\cal O}\left(\eps^0\right).
\eq
Furthermore we have for the process $e^+ e^- \rightarrow \;\mbox{2 jets}$ the additional relation
\bq
 \int\limits_2 d\alpha^{(0,2)}_{(0,1)} & = & \int\limits_1 d\alpha^{(1,1)}_{(0,1)}.
\eq
The correct forms of the unintegrated four-parton antenna functions are listed in the appendix.
The remaining antenna functions can be found in the literature \cite{Gehrmann-DeRidder:2005cm}.

\section{The phase space}
\label{sec:phasespace}

It is a well-known fact, that in the collinear limit spin correlations remain.
For example, the spin-dependent splitting functions for $g \rightarrow g g$
and $g \rightarrow q \bar{q}$ read
\bq
 P^{(0,1)}_{g \rightarrow g g} & = & 
   \frac{2}{s_{ij}}  \left[ - g^{\mu\nu} \left( \frac{2z}{1-z} + \frac{2(1-z)}{z} \right) 
   - 4 (1-\eps) z (1-z) \frac{k^\mu_\perp k^\nu_\perp}{k_\perp^2} \right], \nonumber \\
 P^{(0,1)}_{g \rightarrow q \bar{q}} & = & 
   \frac{2}{s_{ij}} \left[ -g^{\mu\nu} + 4 z (1-z) \frac{k^\mu_\perp k^\nu_\perp}{k_\perp^2} \right],
\eq
where the collinear limit is parameterized as
\bq
\label{collinearlimit}
p_i & = & z p + k_\perp - \frac{k_\perp^2}{z} \frac{n}{2 p n }, \nonumber \\
p_j & = & (1-z)  p - k_\perp - \frac{k_\perp^2}{1-z} \frac{n}{2 p n }.
\eq
Here $n$ is a massless four-vector and the transverse component $k_\perp$ satisfies
$2pk_\perp = 2n k_\perp =0$.
The collinear limits occurs for $k_\perp^2 \rightarrow 0$.
The term
\bq
 {\cal A}_\mu \frac{1}{s_{ij}} \frac{k_\perp^\mu k_\perp^\nu}{k_\perp^2} {\cal A}_\nu
\eq
is proportional to the spin correlation.
In four dimensions the spin-averaged splitting functions are obtained by integrating over the azimuthal angle $\varphi$ 
of $p_i$ around $p$.
By using spin-averaged antenna functions, the subtraction terms have not the same point-wise singular behaviour as the matrix
elements, which is required for  local subtraction terms.
Instead, cancellations of singularities occurs only after an integration over the azimuthal angle 
over all collinear splittings 
of the matrix elements.
For $n$ final-state particles, this is a one-dimensional integration in the $(3n-4)$-dimensional phase space.
It can be shown, that in the single collinear limit, the spin correlation depends on the azimuthal angle $\varphi$
as
\bq
 {\cal A}_\mu \frac{1}{s_{ij}} \frac{k_\perp^\mu k_\perp^\nu}{k_\perp^2} {\cal A}_\nu
 & \sim &
 C_0 + C_2 \cos( 2 \varphi + \alpha).
\eq
One can therefore perform the average with two points, where the azimuthal angle takes the values
\bq
 \varphi, & & \varphi + \frac{\pi}{2},
\eq
while all other coordinates remain fixed.
\\
\\
In detail this is done as follows:
We partition the phase space into different channels.
Within one channel, the phase space is generated iteratively according to
\bq
 d\phi_{n+1} & = & d\phi_n d\phi_{Dipole\;i,j,k}
\eq
For each channel we require that the product $s_{ij} s_{jk}$ is the smallest among all considered channels and that
$s_{ij} < s_{jk}$. Therefore it follows that with channel $(i,j,k)$ also channel $(k,j,i)$ has to be included into
the partioning of the phase space.
For the dipole phase space measure we have
\bq
 d\phi_{dipole}
 & = & \frac{s_{ijk}}{32 \pi^3} 
       \int\limits_0^1 dy \; \left(1-y\right)
       \int\limits_0^1 dz \; 
       \int\limits_0^{2\pi} d\varphi.
\eq
We can therefore generate the $(n+1)$-parton configuration from the $n$-parton configuration by using three random numbers
$u_1$, $u_2$, $u_3$ and by setting
\bq
 y = u_1, \;\;\; z = u_2 \;\;\; \varphi = 2 \pi u_3.
\eq
This defines the invariants as
\bq
 s_{ij} & = & y s_{ijk},
 \nonumber \\
 s_{ik} & = & z (1-y) s_{ijk}, 
 \nonumber \\
 s_{jk} & = & (1-z) (1-y) s_{ijk}.
\eq
From these invariants and the value of $\varphi$ we can reconstruct the four-momenta of the $(n+1)$-parton configuration
\cite{Weinzierl:1999yf}.
The additional phase space weight due to the insertion of the $(n+1)$-th particle is
\bq
 w & = & \frac{s_{ijk}}{16 \pi^2} \left( 1-y \right).
\eq
We have therefore a parametrization of the phase space, such that for every collinear limit the azimuthal average can
be easily performed, while keeping all other coordinates fixed.
It is clear that this procedure can be iterated for multiple collinear emissions.

\section{Numerical results}
\label{sec:num}

As the nominal choice of input parameters I use $N=3$ colors and
$N_f =5$ massless quarks.
I take the electromagnetic coupling to be $\alpha(m_Z) = 1/127.9$ and the strong
coupling to be $\alpha_s(m_Z) = 0.118$. The numerical values of the $Z^0$-mass and width
are $m_Z = 91.187$ GeV and $\Gamma_Z = 2.490$ GeV. For the weak mixing angle I use
$\sin^2 \theta_W = 0.230$.
I take the center of mass energy to be $\sqrt{Q^2} = m_Z$ and I set
the renormalization scale equal to $\mu^2 = Q^2$.

As observable I consider the two-jet cross section. The jets are defined
according to the Durham jet algorithm with $y=0.01$.
The recombination prescription is given by the E-scheme.
The two-jet cross section
has the perturbative expansion
\bq
 \l \sigma \r^{(2-jet)}
 & = & \l \sigma \r^{(0)} \left( 1 + \frac{\alpha_s}{2\pi} B^{(2-jet)}
                               + \left( \frac{\alpha_s}{2\pi} \right)^2 C^{(2-jet)} \right).
\eq
$\l \sigma \r^{(0)}$ is the total hadronic cross section at leading order:
\bq
 \l \sigma \r^{(0)} & = & 40807.4\;\mbox{pb}.
\eq
The NLO and NNLO coefficients $B^{(2-jet)}$ and $C^{(2-jet)}$ have the values
\bq
 B^{(2-jet)} & = & -13.674 \pm 0.004,
 \nonumber \\
 C^{(2-jet)} & = & -231.6 \pm 0.3.
\eq
The new result is the value of the NNLO coefficient $C^{(2-jet)}$, which is obtained directly 
with the methods discussed in this paper.
The correctness of this result can be verified with the help of the known results for the 
total hadronic cross section at NNLO,
the three-jet cross section at NLO and the four-jet cross section at LO. 
We have the perturbative expansions
\bq
 \l \sigma \r^{(tot)}
 & = & \l \sigma \r^{(0)} \left( 1 + \frac{\alpha_s}{2\pi} B^{(tot)}
                               + \left( \frac{\alpha_s}{2\pi} \right)^2 C^{(tot)} \right),
 \nonumber \\
 \l \sigma \r^{(3-jet)}
 & = & \l \sigma \r^{(0)} \left( \frac{\alpha_s}{2\pi} B^{(3-jet)}
                               + \left( \frac{\alpha_s}{2\pi} \right)^2 C^{(3-jet)} \right).
 \nonumber \\
 \l \sigma \r^{(4-jet)}
 & = & \l \sigma \r^{(0)} \left( \frac{\alpha_s}{2\pi} \right)^2 C^{(4-jet)}.
\eq
The perturbative calculation of the inclusive hadronic cross section $\l \sigma \r^{(tot)}$ is actually known
to ${\cal O}(\alpha_s^3)$ \cite{Gorishnii:1991vf,Surguladze:1990tg}, although we need here only the coefficients up to
${\cal O}(\alpha_s^2)$. They are given by \cite{Dine:1979qh,Chetyrkin:1979bj,Celmaster:1979xr}:
\bq
\label{coeff_total_cross_section}
 B^{(tot)} & = & 2,
 \nonumber \\
 C^{(tot)} & = &   \frac{N^2-1}{8 N} \left[ \left( \frac{243}{4} - 44 \zeta_3 \right) N + \frac{3}{4 N} 
                             + \left( 8 \zeta_3 - 11 \right) N_f \right] 
 = 5.64.
\eq
The remaining coefficients are obtained by a simple LO calculation ($B^{(3-jet)}$, $C^{(4-jet)}$) or a 
NLO calculation ($C^{(3-jet)}$).
The values of the coefficients are:
\bq
 B^{(3-jet)} & = & 15.679 \pm 0.004,
 \nonumber \\
 C^{(3-jet)} & = & 153.2 \pm 0.4,
 \nonumber \\
 C^{(4-jet)} & = & 84.39 \pm 0.05.
\eq
These numbers have been obtained with the numerical program presented in this paper. 
The correctness has been checked against already existing programs \cite{Kunszt:1989km,Weinzierl:1999yf}.
Since at order ${\cal O}(\alpha_s)$ any event is either classified as a two-jet or three-jet event, and since
at order ${\cal O}(\alpha_s^2)$ any event is either classified as a two-, three- or four-jet event,
we must have
\bq
 B^{(tot)} & = & B^{(2-jet)} + B^{(3-jet)},
 \nonumber \\
 C^{(tot)} & = & C^{(2-jet)} + C^{(3-jet)} + C^{(4-jet)}.
\eq
We find
\bq
 B^{(2-jet)} + B^{(3-jet)} & = & 2.005 \pm 0.006,
 \nonumber \\
 C^{(2-jet)} + C^{(3-jet)} + C^{(4-jet)}  & = & 6.0 \pm 0.5.
\eq
These numbers agree nicely with the values given in eq.~(\ref{coeff_total_cross_section})
for the total hadronic cross section.

\section{Conclusions}
\label{sec:concl}

In this paper I reported on a numerical program for two-jet observables
in electron-positron annihilation at next-to-next-to-leading order.
To cancel the infra-red divergences the subtraction method with antenna functions is used.
The correctness of the numerical program is verified by comparing the results
for the two-jet cross section to values, which can be obtained indirectly
by subtracting 
from the known result of the total hadronic cross section at $\alpha_s^2$
the next-to-leading order result of the three-jet cross section and the leading
order result of the four-jet cross section.
The numerical program is set up such that -- apart from the specific matrix elements --
nothing is specific to the process $e^+ e^- \rightarrow \mbox{2 jets}$.
I have therefore confidence, that the subtraction method can be extended to other processes
like $e^+ e^- \rightarrow \mbox{3 jets}$ or 
$p p \rightarrow \mbox{2 jets}$.

\subsection*{Acknowledgements}

I would like to thank Th. Gehrmann for providing me with a FORM-file of the four-parton antenna functions.

\begin{appendix}
\section{Antenna functions}
\label{sec:antenna}

Here I list the correct unintegrated four-parton antenna functions
\bq
A_4^{(0)}(q, g, g, \bar{q}), 
 \;\;\;
A_{4,sc}^{(0)}(q, g, g, \bar{q}), 
 \;\;\;
B_4^{(0)}(q, q', \bar{q}', \bar{q}), 
 \;\;\;
C_4^{(0)}(q, q, \bar{q}, \bar{q}).
\eq
The formulae printed in ref.~\cite{Gehrmann-DeRidder:2005cm} differ significantly
from these results.
However, the formulae presented here agree analytically
with a FORM-file obtained from one of the authors of ref.~\cite{Gehrmann-DeRidder:2005cm}.
For the integrated antenna functions I found agreement with the results published 
in ref.~\cite{Gehrmann-DeRidder:2005cm}.
Each antenna is given as a product of a prefactor ${\cal P}$, a symmetry factor ${\cal S}$,
a colour factor ${\cal C}$ and a kinematical factor ${\cal K}$.
The prefactor is given for the unintegrated antenna functions by
\bq
 {\cal P} & = & \left( 8 \pi \alpha_s \right)^2.
\eq
The symmetry and colour factors are all equal to one, except for the antenna $C_4^0$, which has a symmetry factor
of $1/2$.
The kinematical factors are given by
\bq
\lefteqn{
\left.A_4^0(1,2,3,4)\right|_{\cal K}
 = } & & \nonumber \\
 & &
-{\frac {1}{s_{34}\,s_{234}}}+{\frac {4\,s_{12}+4\,s_{34}-8\,s_{1234}}{s_{23}\,s_{123}\,s_{234}}}
+{\frac {2}{s_{12}\,s_{34}}}+{\frac {4}{s_{12}\,s_{234}}}+{\frac {2}{s_{23}\,s_{123}}}+{\frac {2}{s_{23}\,s_{234}}}
-{\frac {2\,{s_{12}}^{2}+2\,{s_{34}}^{2}}{s_{23}\,s_{123}\,s_{234}\,s_{1234}}}
\nonumber \\
& &
+{\frac {-2\,s_{123}-2\,s_{234}+2\,s_{1234}}{s_{12}\,s_{23}\,s_{34}}}+{\frac {{s_{123}}^{2}+{s_{34}}^{2}}{s_{12}\,s_{23}\,s_{234}\,s_{1234}}}
+{\frac {3\,s_{23}-3\,s_{123}}{s_{12}\,s_{234}\,s_{1234}}}+{\frac {3\,s_{34}-3\,s_{234}}{s_{23}\,s_{123}\,s_{1234}}}
+{\frac {4\,s_{12}\,s_{34}}{{s_{23}}^{2}s_{123}\,s_{234}}}
\nonumber \\
& &
+{ \frac {{s_{12}}^{2}+{s_{234}}^{2}}{s_{23}\,s_{123}\,s_{34}\,s_{1234}}}+{\frac {-2\,s_{23}+3\,s_{34}-4\,s_{1234}}{s_{12}\,s_{123}\,s_{234}}}
+{\frac {{s_{123}}^{2}+{s_{234}}^{2}}{s_{12}\,s_{23}\,s_{34}\,s_{1234}}}
+{\frac {2\,s_{23}\,s_{1234}+{s_{23}}^{2}+2\,{s_{1234}}^{2}}{s_{12}\,s_{123}\,s_{34}\,s_{234}}}
\nonumber \\
& &
+{\frac {4}{s_{123}\,s_{34}}}-{\frac {6}{s_{123}\,s_{234}}}+{\frac {6}{s_{123}\,s_{1234}}}+{\frac {6}{s_{234}\,s_{1234}}}
-{\frac {1}{s_{12}\,s_{123}}}+{\frac {3}{{s_{123}}^{2}}}+{\frac {3}{{s_{234}}^{2}}}+{\frac {2}{{s_{23}}^{2}}}
\nonumber \\
& &
-{\frac {-3\,s_{12}+3\,s_{123}}{s_{23}\,s_{234}\,s_{1234}}}-{\frac {{s_{23}}^{2}+{s_{34}}^{2}}{s_{12}\,s_{123}\,s_{234}\,s_{1234}}}
+{\frac {3\,s_{23}-3\,s_{234}}{s_{123}\,s_{34}\,s_{1234}}}-{\frac {-3\,s_{12}+2\,s_{23}+4\,s_{1234}}{s_{123}\,s_{34}\,s_{234}}}
\nonumber \\
& &
-{\frac {2\,s_{23}-s_{234}+2\,s_{1234}}{s_{12}\,s_{123}\,s_{34}}}-{\frac {2\,s_{23}-s_{123}+2\,s_{1234}}{s_{12}\,s_{34}\,s_{234}}}
-{\frac {{s_{12}}^{2}+{s_{23}}^{2}}{s_{123}\,s_{34}\,s_{234}\,s_{1234}}}+{\frac {-2\,s_{12}-2\,s_{234}+2\,s_{1234}}{s_{23}\,s_{123}\,s_{34}}}
\nonumber \\
& &
-{\frac {3\,s_{12}+6\,s_{23}+3\,s_{34}}{s_{123}\,s_{234}\,s_{1234}}}+{\frac {s_{23}}{s_{34}\,{s_{234}}^{2}}}
+{\frac {4\,s_{34}}{s_{23}\,{s_{234}}^{2}}}+{\frac {4\,s_{12}}{s_{23}\,{s_{123}}^{2}}}-{\frac {4\,s_{34}}{{s_{23}}^{2}s_{234}}}
+{\frac {2\,{s_{34}}^{2}}{{s_{23}}^{2}{s_{234}}^{2}}}+{\frac {2\,{s_{12}}^{2}}{{s_{23}}^{2}{s_{123}}^{2}}}
\nonumber \\
& &
-{\frac {4\,s_{12}}{{s_{23}}^{2}s_{123}}}-{\frac {2\,s_{123}+2\,s_{34}-2\,s_{1234}}{s_{12}\,s_{23}\,s_{234}}}
+{\frac {s_{23}}{s_{12}\,{s_{123}}^{2}}}.
\eq
The subleading-colour antenna function $A_{4,sc}^0(1,2,3,4)$ is split into two parts 
\bq
 A_{4,sc}^0(1,2,3,4) & = & \frac{1}{2} a_{4,sc}^0(1,2,3,4) + \frac{1}{2} a_{4,sc}^0(1,3,2,4),
\eq
such that $a_{4,sc}^0(1,2,3,4)$ corresponds to the cyclic ordering $(1,2,3,4)$, while
$a_{4,sc}^0(1,3,2,4)$ corresponds to the cyclic ordering $(1,3,2,4)$. $a_{4,sc}^0(1,2,3,4)$ is given by
\bq
\lefteqn{
\left. a_{4,sc}^0(1,2,3,4)\right|_{\cal K}
 = } & & \nonumber \\
 & &
-{\frac {s_{12}+s_{13}-2\,s_{14}+s_{24}+s_{34}}{{s_{23}}^{2}s_{1234}}}
+{\frac {-s_{12}+s_{23}-s_{13}-s_{14}}{s_{34}\,s_{234}\,s_{1234}}}
-{\frac {s_{13}\,s_{24}}{s_{23}\,s_{123}\,s_{34}\,s_{1234}}}
+{\frac {1}{s_{23}\,s_{1234}}}
\nonumber \\
& &
+{\frac {8\,s_{12}\,s_{14}+3\,s_{12}\,s_{24}+s_{12}\,s_{34}+8\,s_{14}\,s_{13}+s_{13}\,s_{24}+3\,s_{13}\,s_{34}}{2\,s_{23}\,{s_{123}}^{2}s_{1234}}}
\nonumber \\
& &
+{\frac {2\,s_{12}\,s_{13}+4\,s_{12}\,s_{14}+{s_{12}}^{2}+4\,s_{14}\,s_{13}+{s_{13}}^{2}+4\,s_{14}\,s_{24}+4\,s_{14}\,s_{34}+2\,s_{24}\,s_{34}+{s_{24}}^{2}+{s_{34}}^{2}}{2\,s_{23}\,s_{123}\,s_{234}\,s_{1234}}}
\nonumber \\
& &
+{\frac {3\,s_{14}\,s_{13}-3\,s_{14}\,s_{23}+4\,{s_{14}}^{2}-s_{13}\,s_{23}+{s_{13}}^{2}+{s_{23}}^{2}}{s_{12}\,s_{34}\,s_{234}\,s_{1234}}}
+{\frac {s_{23}-2\,s_{13}-2\,s_{14}-s_{24}}{s_{12}\,s_{234}\,s_{1234}}}
\nonumber \\
& &
+{\frac {3\,s_{12}\,s_{24}+s_{12}\,s_{34}+s_{13}\,s_{24}+3\,s_{13}\,s_{34}+8\,s_{14}\,s_{24}+8\,s_{14}\,s_{34}}{2\,s_{23}\,{s_{234}}^{2}s_{1234}}}
+{\frac {-2\,s_{12}-2\,s_{13}-8\,s_{14}+s_{24}+s_{34}}{2\,s_{23}\,s_{234}\,s_{1234}}}
\nonumber \\
& &
+{\frac {s_{13}\,s_{24}}{{s_{23}}^{2}s_{34}\,s_{1234}}}
+{\frac {s_{23}-s_{13}-2\,s_{14}-2\,s_{24}}{s_{123}\,s_{34}\,s_{1234}}}
+{\frac {2\,{s_{14}}^{3}}{s_{123}\,s_{34}\,s_{234}\,s_{1234}\,\left (s_{13}+s_{34}\right )}}
\nonumber \\
& &
+{\frac {2\,{s_{14}}^{3}}{s_{12}\,s_{1234}\,\left (s_{12}+s_{13}\right )\left (s_{12}+s_{24}\right )\left (s_{13}+s_{34}\right )}}
+{\frac {2\,{s_{14}}^{3}}{s_{12}\,s_{34}\,s_{1234}\,\left (s_{12}+s_{24}\right )\left (s_{13}+s_{34}\right )}}
\nonumber \\
& &
+{\frac {2\,{s_{14}}^{3}}{s_{34}\,s_{1234}\,\left (s_{12}+s_{24}\right )\left (s_{13}+s_{34}\right )\left (s_{24}+s_{34}\right )}}
+{\frac {2\,s_{14}}{s_{12}\,s_{1234}\,\left (s_{12}+s_{13}\right )}}
-{\frac {1}{s_{34}\,s_{1234}}}
\nonumber \\
& &
+{\frac {2\,{s_{14}}^{3}}{s_{12}\,s_{123}\,s_{234}\,s_{1234}\,\left (s_{12}+s_{24}\right )}}
+{\frac {2\,s_{14}}{s_{34}\,s_{1234}\,\left (s_{24}+s_{34}\right )}}
-{\frac {1}{s_{12}\,s_{1234}}}
\nonumber \\
& &
-{\frac {-2\,s_{12}\,s_{13}+4\,s_{12}\,s_{14}+s_{12}\,s_{24}-{s_{12}}^{2}+4\,s_{14}\,s_{13}+s_{13}\,s_{34}-{s_{13}}^{2}}{{s_{23}}^{2}s_{123}\,s_{1234}}}
+{\frac {s_{23}\,\left (s_{14}+s_{12}+s_{13}\right )}{s_{34}\,{s_{234}}^{2}s_{1234}}}
\nonumber \\
& &
-{\frac {s_{13}\,s_{24}}{s_{12}\,s_{23}\,s_{234}\,s_{1234}}}
+{\frac {s_{13}\,s_{24}}{s_{12}\,{s_{23}}^{2}s_{1234}}}
+{\frac {4\,s_{14}\,s_{23}+s_{24}\,s_{23}-{s_{23}}^{2}+s_{14}\,s_{24}+s_{14}\,s_{34}}{s_{12}\,s_{123}\,s_{234}\,s_{1234}}}
\nonumber \\
& &
+{\frac {3\,s_{14}\,s_{24}-3\,s_{14}\,s_{23}+4\,{s_{14}}^{2}-s_{24}\,s_{23}+{s_{24}}^{2}+{s_{23}}^{2}}{s_{12}\,s_{123}\,s_{34}\,s_{1234}}}
\nonumber \\
& &
+{\frac {-s_{12}\,s_{24}-s_{13}\,s_{34}-4\,s_{14}\,s_{24}-4\,s_{14}\,s_{34}+2\,s_{24}\,s_{34}+{s_{24}}^{2}+{s_{34}}^{2}}{{s_{23}}^{2}s_{234}\,s_{1234}}}
+{\frac {3\,s_{12}+3\,s_{13}+6\,s_{14}}{2\,{s_{234}}^{2}s_{1234}}}
\nonumber \\
& &
+{\frac {6\,s_{14}+3\,s_{24}+3\,s_{34}}{2\,{s_{123}}^{2}s_{1234}}}
-{\frac {-s_{12}\,s_{14}-s_{13}\,s_{23}-4\,s_{14}\,s_{23}+{s_{23}}^{2}-s_{14}\,s_{13}}{s_{123}\,s_{34}\,s_{234}\,s_{1234}}}
-{\frac {s_{13}\,s_{24}\,\left (s_{12}+s_{13}\right )}{{s_{23}}^{2}s_{123}\,s_{34}\,s_{1234}}}
\nonumber \\
& &
+{\frac {s_{12}-s_{23}+s_{13}+4\,s_{14}+s_{24}+s_{34}}{s_{123}\,s_{234}\,s_{1234}}}
+{\frac {\left (s_{12}+s_{13}\right )\left (2\,s_{12}\,s_{14}+s_{12}\,s_{24}+2\,s_{14}\,s_{13}+s_{13}\,s_{34}\right )}{{s_{23}}^{2}{s_{123}}^{2}s_{1234}}}
\nonumber \\
& &
+{\frac {6\,s_{14}+3\,s_{13}+3\,s_{24}}{s_{12}\,s_{34}\,s_{1234}}}
-{\frac {s_{13}\,s_{24}\,\left (s_{24}+s_{34}\right )}{s_{12}\,{s_{23}}^{2}s_{234}\,s_{1234}}}
+{\frac {2\,s_{14}\,\left (s_{12}+s_{13}\right )\left (s_{24}+s_{34}\right )}{{s_{23}}^{2}s_{123}\,s_{234}\,s_{1234}}}
\nonumber \\
& &
+{\frac {s_{14}\,\left (s_{12}+s_{23}+2\,s_{14}\right )}{s_{34}\,s_{234}\,s_{1234}\,\left (s_{13}+s_{34}\right )}}
+{\frac {\left (s_{24}+s_{34}\right )\left (s_{13}\,s_{34}+2\,s_{14}\,s_{34}+s_{12}\,s_{24}+2\,s_{14}\,s_{24}\right )}{{s_{23}}^{2}{s_{234}}^{2}s_{1234}}}
\nonumber \\
& &
-{\frac {-s_{12}-s_{13}+8\,s_{14}+2\,s_{24}+2\,s_{34}}{2\,s_{23}\,s_{123}\,s_{1234}}}
+{\frac {s_{14}\,\left (s_{12}+2\,s_{14}\right )}{s_{34}\,s_{1234}\,\left (s_{13}+s_{34}\right )\left (s_{24}+s_{34}\right )}}
+{\frac {s_{14}\,\left (s_{34}+s_{23}+2\,s_{14}\right )}{s_{12}\,s_{123}\,s_{1234}\,\left (s_{12}+s_{24}\right )}}
\nonumber \\
& &
+{\frac {s_{14}\,\left (s_{13}+s_{23}+2\,s_{14}\right )}{s_{12}\,s_{234}\,s_{1234}\,\left (s_{12}+s_{24}\right )}}
+{\frac {s_{14}\,\left (s_{23}+s_{24}+2\,s_{14}\right )}{s_{123}\,s_{34}\,s_{1234}\,\left (s_{13}+s_{34}\right )}}
+{\frac {s_{14}\,\left (2\,s_{14}+s_{24}\right )}{s_{12}\,s_{34}\,s_{1234}\,\left (s_{13}+s_{34}\right )}}
\nonumber \\
& &
+{\frac {s_{14}\,\left (s_{13}+2\,s_{14}\right )}{s_{34}\,s_{1234}\,\left (s_{12}+s_{24}\right )\left (s_{24}+s_{34}\right )}}
+{\frac {s_{14}\,\left (s_{13}+2\,s_{14}\right )}{s_{12}\,s_{34}\,s_{1234}\,\left (s_{12}+s_{24}\right )}}
+{\frac {s_{14}\,\left (2\,s_{14}+s_{24}\right )}{s_{12}\,s_{1234}\,\left (s_{12}+s_{13}\right )\left (s_{13}+s_{34}\right )}}
\nonumber \\
& &
+{\frac {s_{14}\,\left (s_{34}+2\,s_{14}\right )}{s_{12}\,s_{1234}\,\left (s_{12}+s_{13}\right )\left (s_{12}+s_{24}\right )}}
-{\frac {\left (s_{23}-s_{14}\right )\left ({s_{23}}^{2}-2\,s_{14}\,s_{23}+2\,{s_{14}}^{2}\right )}{s_{12}\,s_{123}\,s_{34}\,s_{234}\,s_{1234}}}
+{\frac {s_{23}\,\left (s_{14}+s_{24}+s_{34}\right )}{s_{12}\,{s_{123}}^{2}s_{1234}}}
\nonumber \\
& &
-{\frac {-s_{23}+s_{14}+s_{24}+s_{34}}{s_{1234}\,s_{123}\,s_{12}}},
\eq
Finally the antenna functions with four quarks:
\bq
\lefteqn{
\left. B_4^0(1,2,3,4)\right|_{\cal K}
 = } & & \nonumber \\
 & &
-{\frac {-2\,s_{14}\,s_{24}\,s_{34}-s_{12}\,s_{24}\,s_{34}-s_{13}\,s_{24}\,s_{34}+s_{13}\,{s_{24}}^{2}+s_{12}\,{s_{34}}^{2}}{{s_{23}}^{2}{s_{234}}^{2}s_{1234}}}
\nonumber \\
& &
- \frac{1}{{s_{23}}^{2}s_{234}\,s_{1234}\,s_{123}} 
 \left[2\,s_{13}\,s_{14}\,s_{24}+2\,s_{12}\,s_{14}\,s_{34}-s_{12}\,s_{13}\,s_{24}-s_{12}\,s_{24}\,s_{34}-s_{13}\,s_{24}\,s_{34}+{s_{13}}^{2}s_{24}
 \right. \nonumber \\
 & & \left.
+s_{13}\,{s_{24}}^{2}-s_{12}\,s_{13}\,s_{34}+s_{12}\,{s_{34}}^{2}+{s_{12}}^{2}s_{34}\right]
\nonumber \\
& &
-{\frac {-2\,s_{12}\,s_{13}\,s_{14}-s_{12}\,s_{13}\,s_{24}+{s_{13}}^{2}s_{24}-s_{12}\,s_{13}\,s_{34}+{s_{12}}^{2}s_{34}}{{s_{23}}^{2}s_{1234}\,{s_{123}}^{2}}}
+{\frac {s_{14}\,s_{24}+s_{14}\,s_{34}+s_{12}\,s_{24}+s_{13}\,s_{34}}{s_{23}\,{s_{234}}^{2}s_{1234}}}
\nonumber \\
& &
+{\frac {s_{14}\,\left (s_{24}+s_{12}+s_{13}+s_{34}+2\,s_{14}\right )}{s_{23}\,s_{234}\,s_{1234}\,s_{123}}}
+{\frac {s_{12}\,s_{14}+s_{14}\,s_{13}+s_{12}\,s_{24}+s_{13}\,s_{34}}{s_{23}\,s_{1234}\,{s_{123}}^{2}}}
+{\frac {2\,s_{14}}{s_{234}\,s_{1234}\,s_{123}}},
\eq
\bq
\lefteqn{
\left. C_4^0(1,2,3,4)\right|_{\cal K}
 = } & & \nonumber \\
 & &
-{\frac {s_{12}\,s_{14}\,s_{13}}{s_{24}\,s_{124}\,s_{23}\,s_{1234}\,s_{123}}}
+{\frac {s_{12}\,\left (s_{14}-s_{13}\right )}{2\,s_{24}\,s_{124}\,s_{23}\,s_{1234}}}
+{\frac {-s_{12}\,s_{14}-s_{14}\,s_{13}-s_{14}\,s_{23}-{s_{14}}^{2}+s_{12}\,s_{13}+s_{12}\,s_{34}}{2\,s_{24}\,s_{124}\,s_{234}\,s_{1234}}}
\nonumber \\
& &
-{\frac {s_{12}\,s_{14}}{s_{24}\,s_{124}\,s_{1234}\,s_{123}}}
-{\frac {s_{12}\,\left (s_{14}-s_{13}\right )}{2\,s_{24}\,s_{23}\,s_{1234}\,s_{123}}}
-{\frac {s_{12}}{s_{24}\,s_{23}\,s_{1234}}}
+{\frac {s_{34}\,\left (s_{14}+s_{12}+s_{13}\right )}{s_{24}\,{s_{234}}^{2}s_{1234}}}
\nonumber \\
& &
+{\frac {s_{12}\,s_{14}+s_{14}\,s_{13}+s_{14}\,s_{23}+{s_{14}}^{2}-s_{12}\,s_{13}+s_{12}\,s_{34}}{2\,s_{24}\,s_{234}\,s_{1234}\,s_{123}}}
+{\frac {s_{12}}{s_{24}\,s_{234}\,s_{1234}}}
\nonumber \\
& &
-{\frac {s_{12}\,s_{14}-s_{14}\,s_{13}+s_{12}\,s_{24}-s_{13}\,s_{24}-s_{12}\,s_{13}-{s_{13}}^{2}}{2\,s_{124}\,s_{23}\,s_{234}\,s_{1234}}}
-{\frac {s_{12}\,s_{13}}{s_{124}\,s_{23}\,s_{1234}\,s_{123}}}
+{\frac {s_{12}}{2\,s_{124}\,s_{23}\,s_{1234}}}
\nonumber \\
& &
-{\frac {s_{14}+s_{12}-s_{13}}{2\,s_{124}\,s_{234}\,s_{1234}}}
-{\frac {s_{12}}{s_{124}\,s_{1234}\,s_{123}}}
-{\frac {s_{24}\,\left (s_{14}+s_{12}+s_{13}\right )}{s_{23}\,{s_{234}}^{2}s_{1234}}}
\nonumber \\
& &
-{\frac {-s_{12}\,s_{14}+s_{14}\,s_{13}+s_{12}\,s_{24}+s_{13}\,s_{24}+s_{12}\,s_{13}+{s_{13}}^{2}}{2\,s_{23}\,s_{234}\,s_{1234}\,s_{123}}}
+{\frac {s_{14}+2\,s_{12}+s_{13}}{s_{23}\,s_{234}\,s_{1234}}}
+{\frac {s_{12}}{2\,s_{23}\,s_{1234}\,s_{123}}}
\nonumber \\
& &
-{\frac {s_{14}+s_{12}+s_{13}}{{s_{234}}^{2}s_{1234}}}
+{\frac {s_{14}-s_{12}-s_{13}}{2\,s_{234}\,s_{1234}\,s_{123}}}.
\eq

\end{appendix}



\begin{thebibliography}{10}

\bibitem{Tarasov:1996br}
O.~V. Tarasov,
\newblock Phys. Rev. {\bf D54}, 6479 (1996), hep-th/9606018;
\newblock Nucl. Phys. {\bf B502}, 455 (1997), hep-ph/9703319.

\bibitem{Smirnov:1999gc}
V.~A. Smirnov,
\newblock Phys. Lett. {\bf B460}, 397 (1999), hep-ph/9905323;
\newblock Phys. Lett. {\bf B491}, 130 (2000), hep-ph/0007032;
\newblock Phys. Lett. {\bf B500}, 330 (2001), hep-ph/0011056.

\bibitem{Tausk:1999vh}
J.~B. Tausk,
\newblock Phys. Lett. {\bf B469}, 225 (1999), hep-ph/9909506.

\bibitem{Laporta:2000dc}
S.~Laporta,
\newblock Phys. Lett. {\bf B504}, 188 (2001), hep-ph/0102032;
\newblock Int. J. Mod. Phys. {\bf A15}, 5087 (2000), hep-ph/0102033.

\bibitem{Gehrmann:1999as}
T.~Gehrmann and E.~Remiddi,
\newblock Nucl. Phys. {\bf B580}, 485 (2000), hep-ph/9912329;
\newblock Nucl. Phys. {\bf B601}, 248 (2001), hep-ph/0008287;
\newblock Nucl. Phys. {\bf B601}, 287 (2001), hep-ph/0101124.

\bibitem{Moch:2001zr}
S.~Moch, P.~Uwer, and S.~Weinzierl,
\newblock J. Math. Phys. {\bf 43}, 3363 (2002), hep-ph/0110083.

\bibitem{Weinzierl:2002hv}
S.~Weinzierl,
\newblock Comput. Phys. Commun. {\bf 145}, 357 (2002), math-ph/0201011.

\bibitem{Moch:2005uc}
S.~Moch and P.~Uwer,
\newblock Comput. Phys. Commun. {\bf 174}, 759 (2006), math-ph/0508008.

\bibitem{Czakon:2005rk}
M.~Czakon,
\newblock (2005), hep-ph/0511200.

\bibitem{Anastasiou:2004vj}
C.~Anastasiou and A.~Lazopoulos,
\newblock JHEP {\bf 07}, 046 (2004), hep-ph/0404258.

\bibitem{Bern:2000ie}
Z.~Bern, L.~Dixon, and A.~Ghinculov,
\newblock Phys. Rev. {\bf D63}, 053007 (2001), hep-ph/0010075.

\bibitem{Bern:2000dn}
Z.~Bern, L.~Dixon, and D.~A. Kosower,
\newblock JHEP {\bf 01}, 027 (2000), hep-ph/0001001.

\bibitem{Anastasiou:2000kg}
C.~Anastasiou, E.~W.~N. Glover, C.~Oleari, and M.~E. Tejeda-Yeomans,
\newblock Nucl. Phys. {\bf B601}, 318 (2001), hep-ph/0010212;
\newblock Nucl. Phys. {\bf B601}, 341 (2001), hep-ph/0011094;
\newblock Phys. Lett. {\bf B506}, 59 (2001), hep-ph/0012007;
\newblock Nucl. Phys. {\bf B605}, 486 (2001), hep-ph/0101304.

\bibitem{Glover:2001af}
E.~W.~N. Glover, C.~Oleari, and M.~E. Tejeda-Yeomans,
\newblock Nucl. Phys. {\bf B605}, 467 (2001), hep-ph/0102201.

\bibitem{Bern:2001dg}
Z.~Bern, A.~De~Freitas, L.~J. Dixon, A.~Ghinculov, and H.~L. Wong,
\newblock JHEP {\bf 11}, 031 (2001), hep-ph/0109079.

\bibitem{Bern:2001df}
Z.~Bern, A.~De~Freitas, and L.~J. Dixon,
\newblock JHEP {\bf 09}, 037 (2001), hep-ph/0109078;
\newblock JHEP {\bf 03}, 018 (2002), hep-ph/0201161.

\bibitem{Garland:2001tf}
L.~W. Garland, T.~Gehrmann, E.~W.~N. Glover, A.~Koukoutsakis, and E.~Remiddi,
\newblock Nucl. Phys. {\bf B627}, 107 (2002), hep-ph/0112081;
\newblock Nucl. Phys. {\bf B642}, 227 (2002), hep-ph/0206067.

\bibitem{Moch:2002hm}
S.~Moch, P.~Uwer, and S.~Weinzierl,
\newblock Phys. Rev. {\bf D66}, 114001 (2002), hep-ph/0207043.

\bibitem{Kosower:2002su}
D.~A. Kosower,
\newblock Phys. Rev. {\bf D67}, 116003 (2003), hep-ph/0212097;
\newblock Phys. Rev. Lett. {\bf 91}, 061602 (2003), hep-ph/0301069;
\newblock Phys. Rev. {\bf D71}, 045016 (2005), hep-ph/0311272.

\bibitem{Weinzierl:2003fx}
S.~Weinzierl,
\newblock JHEP {\bf 03}, 062 (2003), hep-ph/0302180;
\newblock JHEP {\bf 07}, 052 (2003), hep-ph/0306248.

\bibitem{Anastasiou:2003gr}
C.~Anastasiou, K.~Melnikov, and F.~Petriello,
\newblock Phys. Rev. {\bf D69}, 076010 (2004), hep-ph/0311311.

\bibitem{Gehrmann-DeRidder:2003bm}
A.~Gehrmann-De~Ridder, T.~Gehrmann, and G.~Heinrich,
\newblock Nucl. Phys. {\bf B682}, 265 (2004), hep-ph/0311276.

\bibitem{Gehrmann-DeRidder:2004tv}
A.~Gehrmann-De~Ridder, T.~Gehrmann, and E.~W.~N. Glover,
\newblock Nucl. Phys. {\bf B691}, 195 (2004), hep-ph/0403057;
\newblock Phys. Lett. {\bf B612}, 36 (2005), hep-ph/0501291;
\newblock Phys. Lett. {\bf B612}, 49 (2005), hep-ph/0502110.

\bibitem{Gehrmann-DeRidder:2005cm}
A.~Gehrmann-De~Ridder, T.~Gehrmann, and E.~W.~N. Glover,
\newblock JHEP {\bf 09}, 056 (2005), hep-ph/0505111.

\bibitem{Binoth:2004jv}
T.~Binoth and G.~Heinrich,
\newblock Nucl. Phys. {\bf B693}, 134 (2004), hep-ph/0402265.

\bibitem{Heinrich:2006sw}
G.~Heinrich,
\newblock (2006), hep-ph/0601062.

\bibitem{Kilgore:2004ty}
W.~B. Kilgore,
\newblock Phys. Rev. {\bf D70}, 031501 (2004), hep-ph/0403128.

\bibitem{Frixione:2004is}
S.~Frixione and M.~Grazzini,
\newblock JHEP {\bf 06}, 010 (2005), hep-ph/0411399.

\bibitem{Somogyi:2005xz}
G.~Somogyi, Z.~Trocsanyi, and V.~Del~Duca,
\newblock JHEP {\bf 06}, 024 (2005), hep-ph/0502226.

\bibitem{Frixione:1996ms}
S.~Frixione, Z.~Kunszt, and A.~Signer,
\newblock Nucl. Phys. {\bf B467}, 399 (1996), hep-ph/9512328.

\bibitem{Catani:1997vz}
S.~Catani and M.~H. Seymour,
\newblock Nucl. Phys. {\bf B485}, 291 (1997), hep-ph/9605323.

\bibitem{Dittmaier:1999mb}
S.~Dittmaier,
\newblock Nucl. Phys. {\bf B565}, 69 (2000), hep-ph/9904440.

\bibitem{Phaf:2001gc}
L.~Phaf and S.~Weinzierl,
\newblock JHEP {\bf 04}, 006 (2001), hep-ph/0102207.

\bibitem{Catani:2002hc}
S.~Catani, S.~Dittmaier, M.~H. Seymour, and Z.~Trocsanyi,
\newblock Nucl. Phys. {\bf B627}, 189 (2002), hep-ph/0201036.

\bibitem{Hepp:1966eg}
K.~Hepp,
\newblock Commun. Math. Phys. {\bf 2}, 301 (1966).

\bibitem{Roth:1996pd}
M.~Roth and A.~Denner,
\newblock Nucl. Phys. {\bf B479}, 495 (1996), hep-ph/9605420.

\bibitem{Binoth:2000ps}
T.~Binoth and G.~Heinrich,
\newblock Nucl. Phys. {\bf B585}, 741 (2000), hep-ph/0004013.

\bibitem{Anastasiou:2004qd}
C.~Anastasiou, K.~Melnikov, and F.~Petriello,
\newblock Phys. Rev. Lett. {\bf 93}, 032002 (2004), hep-ph/0402280;
\newblock Phys. Rev. Lett. {\bf 93}, 262002 (2004), hep-ph/0409088;
\newblock Nucl. Phys. {\bf B724}, 197 (2005), hep-ph/0501130.

\bibitem{Matsuura:1988wt}
T.~Matsuura and W.~L. van Neerven,
\newblock Z. Phys. {\bf C38}, 623 (1988).

\bibitem{Matsuura:1989sm}
T.~Matsuura, S.~C. van~der Marck, and W.~L. van Neerven,
\newblock Nucl. Phys. {\bf B319}, 570 (1989).

\bibitem{Kramer:1987sg}
G.~Kramer and B.~Lampe,
\newblock Z. Phys. {\bf C34}, 497 (1987);
\newblock Z. Phys. {\bf C42}, 504 (1989),
\newblock Erratum.

\bibitem{Ellis:1981wv}
R.~K. Ellis, D.~A. Ross, and A.~E. Terrano,
\newblock Nucl. Phys. {\bf B178}, 421 (1981).

\bibitem{Schuler:1987ej}
G.~A. Schuler, S.~Sakakibara, and J.~G. K\"orner,
\newblock Phys. Lett. {\bf B194}, 125 (1987).

\bibitem{Korner:1990sj}
J.~G. K\"orner and P.~Sieben,
\newblock Nucl. Phys. {\bf B363}, 65 (1991).

\bibitem{Giele:1992vf}
W.~T. Giele and E.~W.~N. Glover,
\newblock Phys. Rev. {\bf D46}, 1980 (1992).

\bibitem{Ali:1979rz}
A.~Ali {\em et~al.},
\newblock Phys. Lett. {\bf B82}, 285 (1979);
\newblock Nucl. Phys. {\bf B167}, 454 (1980).

\bibitem{Kosower:1998zr}
D.~A. Kosower,
\newblock Phys. Rev. {\bf D57}, 5410 (1998), hep-ph/9710213.

\bibitem{Weinzierl:1999yf}
S.~Weinzierl and D.~A. Kosower,
\newblock Phys. Rev. {\bf D60}, 054028 (1999), hep-ph/9901277.

\bibitem{Gorishnii:1991vf}
S.~G. Gorishnii, A.~L. Kataev, and S.~A. Larin,
\newblock Phys. Lett. {\bf B259}, 144 (1991).

\bibitem{Surguladze:1990tg}
L.~R. Surguladze and M.~A. Samuel,
\newblock Phys. Rev. Lett. {\bf 66}, 560 (1991).

\bibitem{Dine:1979qh}
M.~Dine and J.~R. Sapirstein,
\newblock Phys. Rev. Lett. {\bf 43}, 668 (1979).

\bibitem{Chetyrkin:1979bj}
K.~G. Chetyrkin, A.~L. Kataev, and F.~V. Tkachov,
\newblock Phys. Lett. {\bf B85}, 277 (1979).

\bibitem{Celmaster:1979xr}
W.~Celmaster and R.~J. Gonsalves,
\newblock Phys. Rev. Lett. {\bf 44}, 560 (1980).

\bibitem{Kunszt:1989km}
Z.~Kunszt, P.~Nason, G.~Marchesini, and B.~R. Webber,
\newblock ``QCD at LEP'', Proceedings of the 1989 LEP Physics Workshop,
  Geneva, Switzerland, Feb 20, 1989.

\end{thebibliography}
\end{document}